\documentclass[showpacs,aps]{revtex4}
\usepackage{eurosym}
\usepackage{graphicx}
\usepackage{amsmath,amssymb}
\usepackage{color}

\def\bc{\begin{center}}
\def\ec{\end{center}}

\def\beq{\begin{equation}}
\def\eeq{\end{equation}}

\begin{document}

\title{Inelastic processes in Na$^{+}-$Ne, Ar and Ne$^{+},$ Ar$^{+}-$Na
collisions in energy range $0.5-14$ keV}
\author{R. A. Lomsadze$^{1}$, M. R. Gochitashvili$^{1}$, R. Ya. Kezerashvili$%
^{2,3}$}
\affiliation{$^{1}$Tbilisi State University, Tbilisi, 0128, Georgia\\
$^{2}$New York City College of Technology, The City University of New York,
Brooklyn, NY 11201, USA \\
$^{3}$The Graduate School and University Center, The City University of New
York, New York, NY 10016, USA }
\date{\today}

\begin{abstract}
Absolute cross sections for charge-exchange, ionization and excitation in Na$%
^{+}-$Ne and Na$^{+}-$Ar collisions were measured in the ion energy range $%
0.5-10$ keV using a refined version of a capacitor method, and collision and
optical spectroscopy methods simultaneously in the same experimental set-up.
Ionization cross sections for Ne$^{+}-$Na and Ar$^{+}-$Na collisions are
measured at the energies of $2-14$ keV using a crossed-beam spectroscopy
method. The experimental data and the schematic correlation diagrams are
used to analyze and determine the mechanisms for these processes. For the
charge-exchange process in Na$^{+}$ $-$Ar collisions two nonadiabatic
regions are revealed and mechanisms responsible for these regions are
explained. Structural peculiarity on the excitation function for the
resonance lines of argon atoms in Na$^{+}$ $-$Ar collisions are observed and
the possible mechanisms of this phenomenon are explored. The measured
ionization cross sections for Na$^{+}-$Ne and Ne$^{+}-$Na collisions in
conjunction with the Landau-Zener formula are used to determine the coupling
matrix element and transition probability in a region of pseudo-crossing of
the potential curves.
\end{abstract}

\maketitle

\section{INTRODUCTION}

Studies of inelastic processes in slow ion--atom collisions yield extensive
data in a quasimolecular mechanism of interaction between the colliding
particles. Molecular terms of the system of colliding particles are used in
the description of inelastic processes in ion--atom collisions. At present,
such terms have been calculated for only a small number of simple systems,
so that schematic correlation diagrams for molecular orbitals are widely
used but provide only a qualitative explanation of the known features of
these processes. However, theoretical calculations alone are insufficient
for the complete understanding of the interaction picture due to complicated
many--channel character of the processes under investigation. Moreover,
because different mechanisms are active at various internuclear distances
their study requires several experimental techniques all working together to
make a consistent interpretation possible.

On the experimental side, for investigation of the molecular potential
curves of a scattering system a set of differential cross sections for
various channels following these potential curves are needed. However, due
to the limited energy resolution of the differential scattering technique
most often is problematic an identification of the specific final states,
and only possible a determination of two groups of levels corresponding to
one-- and two--electron excitation. Thus, in order to infer the differential
scattering cross section which results when the system follows specified
molecular potential curves on the incoming and outgoing portions of the
trajectory, differential energy--loss spectra must be supplemented by cross
sections for a photon emission and electron ejection as a function of the
respective energies.

Despite many experimental studies of the alkali ion--gas collisions, which
have been carried out by various experimental methods \cite{Mouzon9, Moe5,
Flaks1, Ogurtsov4, Matveev7, Matveev8, Afrosimov2, Francois13, Lorents12,
Latypov11, Kita18, V. V. Afrosimov16 21, Bidin6, Hegerberg3, Jrgensen20,
Os22, Kita17, Kita2000, Lomsadze23, Kita10, KitaR, Kez25, Kita19} available data for
the absolute cross section for the most inelastic processes are
contradictory \cite{Mouzon9, Moe5, Flaks1, Latypov11, Bidin6} and in some
cases unreliable \cite{Ogurtsov4}.

The inelastic collision mechanism has been studied experimentally and
theoretically for the systems Li$^{+}$--He and Li$^{+}$--Ne in Refs. \cite%
{Francois13, Lorents12, Kita2000, Junker, Sidis}. Double differential cross
sections have also been measured in Ref. \cite{V. V. Afrosimov16 21}.

The results of the measurements of the excitation function for Na$^{+}$--He
and K$^{+}$--He colliding pairs in arbitrary units are reported in Refs. 
\cite{Matveev7} and \cite{Matveev8}. A relative differential cross section
for Na$^{+}$--Ar is measured in Ref. \cite{Kita17}. The absolute values of
the differential cross section at two fixed energies $E=200$ eV and $E=350$
eV of Na$^{+}$ ions were determined by using the experimental integral cross
sections and the repulsive potential deduced also experimentally are
reported in Ref. \cite{Kita18}. Excitation processes in Cs$^{+}$--Ar
collisions were studied at laboratory collision energies of 0.2--1 keV by
means of differential scattering spectroscopy in Ref.\cite{KitaR}. Recently
in Ref. \cite{Kita19}\ a comprehensive study of excitation mechanisms in Na$%
^{+}$--He and K$^{+}$--He collisions at ions energies range of 1.0--1.5 keV
by differential scattering spectroscopy was reported. Double differential
cross sections were measured by detecting all scattering particles (Na$^{+}$%
, Na, K$^{+}$, K, He$^{+}$ and He) over a wide range of center-of-mass
scattering angles. A systematic study of inelastic processes in K$^{+}$--He
collisions is presented in Ref. \cite{Kez25}. Absolute cross sections for
charge exchange, ionization, stripping, and excitation in K$^{+}$--He
collisions were measured in the ion energy range 0.7--10 keV.

For Na$^{+}$--Ar collisions an energy spectrum of electrons ejected from
autoionizing states of Ar atoms at $E=15$ keV \cite{Jrgensen20} have been
briefly reported, but no results exist for lower energy collisions.

The most comprehensive approach to study ion-atom collisions with closed
electron shells has so far been carried out only for a Na$^{+}$--Ne pair 
\cite{Os22}. Though, measurements were performed at a limited energy
interval. Therefore, in our study attention will be focused on the extension
of the energy interval from 0.7 to 10 keV.

In earlier publication \cite{Lomsadze23} we have reported limited results
for inelastic processes realized in Na$^{+}$--Ar collisions. In the present
study, the differential cross section as well as the energy loss spectrum
for Na$^{+}$--Ar collisions will be investigated additionally over a wide
range of energy and scattering angles.

To our mind the reason for a lack of systematic measurements and reliable
data for alkali metal ions and rare gas atoms collisions, and having not at
all measurements for inert gas ions collisions with alkali metal atoms are
linked with experimental difficulties. Mainly, in the case of the alkali
metal ions and rare gas atoms collisions these difficulties related to
collections and detections of secondary particles, while, in case of the
rare gas ions and alkali metal atoms collisions mostly related to the
preparation of alkali metal atoms as a target.

The lack of systematic absolute cross sections measurements for Na$^{+}-$Ne
and Ne$^{+}-$Na colliding pairs motivated the present detailed investigation
of the primary mechanisms for these collision processes.

The collision of a Na$^{+}$ ion beam with Ne and Ar atoms leads mainly to
the following processes:

\begin{eqnarray}
\text{Na}^{\text{+}}+\text{Ne} &\rightarrow &\text{Na}+\text{Ne}^{+},
\label{Ne1} \\
&\rightarrow &\text{Na}^{+}+\text{Ne}^{+}+e,  \label{Ne2} \\
&\rightarrow &\text{Na}^{+}+\text{Ne}^{\ast },  \label{Ne3}
\end{eqnarray}

\bigskip and%
\begin{eqnarray}
\text{Na}^{\text{+}}+\text{Ar} &\rightarrow &\text{Na}+\text{Ar}^{+},
\label{Ar1} \\
&\rightarrow &\text{Na}^{+}+\text{Ar}^{+}+e,  \label{Ar2} \\
&\rightarrow &\text{Na}^{+}+\text{Ar}^{\ast }.  \label{Ar3}
\end{eqnarray}

\bigskip In the charge-exchange processes (\ref{Ne1}) and (\ref{Ar1}) the Na
atom and Ne$^{+}$ or Ar$^{+}$ ion can be in the ground states or in
different excited states. The reactions (\ref{Ne2}) and (\ref{Ar2})
represent the ionization processes for the target atoms, that include
different channels for the excitation of the Na$^{+}$ ion or/and Ne$^{+}($Ar$%
^{+})$ ion, as well as the excitation of autoionization states of the target
atom that leads to its ionization. The excitation processes (\ref{Ne3}) and (%
\ref{Ar3}) include different channels for excitation of the Na$^{+}$ ion
or/and Ne(Ar) atom.

In collisions of Ne$^{+}$ and Ar$^{+}$ ions with a Na atom let's consider
only charge-exchange and ionization processes: 
\begin{eqnarray}
\text{Ne}^{\text{+}}+\text{Na} &\rightarrow &\text{Ne}+\text{Na}^{+},
\label{NeNa1} \\
&\rightarrow &\text{Ne}^{+}+\text{Na}^{+}+e  \label{NeNa2}
\end{eqnarray}

\begin{eqnarray}
\text{Ar}^{\text{+}}+\text{Na} &\rightarrow &\text{Ar}+\text{Na}^{+},
\label{ArNa1} \\
&\rightarrow &\text{Ar}^{+}+\text{Na}^{+}+e,  \label{ArNa2}
\end{eqnarray}

The processes (\ref{NeNa1}) and (\ref{ArNa1}) represent the charge-exchange
reactions when the Ne(Ar) atom and Na$^{+}$ ion can be in the ground state
or in different excited states, while a result of the ionization reactions (%
\ref{NeNa2}) and (\ref{ArNa2}) includes different channels for the
excitation of the Na$^{+}$ ion or/and Ne$^{+}($Ar$^{+})$ ion, as well as
excitation of autoionization states of the target atom that leads to its
ionization.

The simultaneous study of the processes \ (\ref{Ne1}) and (\ref{NeNa1}), as
well as \ (\ref{Ar1}) and (\ref{ArNa1}), when products of the reactions are
in the ground states represent a fundamental interest because they are the
reversible processes. A symmetry of these processes arises from the concept
of time reversal. The symmetry transformation that changes a physical system
with a given sense of the time evolution into another with the opposite
sense is called time reversal. The symmetry of these processes means that
the probabilities for the processes Na$^{+}+$Ne$\rightarrow $Na$(g.s.)$+Ne$%
^{+}(g.s.)$ ($g.s.$ hereafter refers as the ground state) and Na$^{\text{+}%
}+ $Ar$\rightarrow $Na$(g.s.)+$Ar$^{+}(g.s.)$ are the same as the
probabilities for the reversed processes Ne$^{+}+$Na$\rightarrow $Ne$(g.s.)$%
+Na$^{+}(g.s)$ and Ar$^{\text{+}}+$Na$\rightarrow $Ar$(g.s.)+$Na$^{+}(g.s.).$
This reciprocity leads to the important principle of detailed balance that
relates the cross section, for example, for the reaction Na$^{+}$+Ne$%
\rightarrow $Na$(g.s.)+$Ne$^{+}(g.s.)$ with that of the time-reversed
reaction Ne$^{+}$+Na$\rightarrow $Ne$(g.s)$+Na$^{+}(g.s.).$ Motivated by the 
$CP$ violation found in the neutral kaon system \cite{CP}, several tests of
time reversal invariance in low-energy nuclear physics have been performed
in the weak, electromagnetic, and strong interactions and were consistent
with the time irreversibility \cite{Annu.Rev}. Since an experimental test of
the detailed balance and time irreversibility in the reactions $^{27}$AI$+p$ 
$\leftrightarrows ^{24}$Mg$+\alpha $ \cite{Blanke} this problem still
remains essential and important \cite{Uzikov}. The detailed balance and time
irreversibility have been also studied in kinetics \cite{Alberty}. In our
best knowledge there is no studies of time irreversibility and detailed
balance in atomic collisions. The above mentioned processes are good
candidates for the such test. However, experimental difficulties of
identifications of the ground states of the charge-exchange products do not
allow us to proceed accurate measurements. Though this task poses
challenges, they are challenges we are ready to face.

A simultaneous study of the ionization processes \ (\ref{Ne2}) and (\ref%
{NeNa2}), as well as (\ref{Ar2}) and (\ref{ArNa2}) also represent a
particular interest. A close attention to the reactions (\ref{Ne2}) and (\ref%
{NeNa2}), and (\ref{Ar2}) and (\ref{ArNa2}) shows the similarity of the
reactions products. One of the objectives of this work is to show that in
some cases (e.g. for Na$^{+}-$Ne) the information extracted from the
theoretical calculations can be obtained experimentally using a simple
approach, namely by measuring a relative ionization cross section for Na$%
^{+}-$Ne and Ne$^{+}-$Na colliding pairs.

Below we report the absolute total and differential cross sections for
charge-exchange, ionization as well as the excitation of the both the
projectile and target particles and energy loss spectra in collisions of Na$%
^{+}$ ions with Ne and Ar atoms, and Ne$^{+}$and Ar$^{+}$ ions with a Na
atom. In later case, the cross section of ionization will be reported for
the first time.

The remainder of this paper is organized in the following way. In Sec. II
the experimental set-ups and procedures are described and four different
experimental methods of measurements for collision experiments are
presented. Here we introduce the procedures for measurements of the absolute
total and differential cross sections for charge-exchange, ionization and
excitation. Results of measurements for the processes \ (\ref{Ne1})--(\ref%
{Ne3}), (\ref{Ar1})--(\ref{Ar3}), (\ref{NeNa2}), and (\ref{ArNa2}), the
comparison of our measurements with results of previous experimental
studies, and discussion of mechanisms for different processes occurring in Na%
$^{+}$--Ar collision are given in Sec.III. The method and procedure for
estimation of some theoretical parameters (transition probability, matrix
element) from our experimental results are reported in Sec. IV. Finally, in
Sec.V we summarize our investigations and present the conclusions.

\section{EXPERIMENTAL SET-UPS AND PROCEDURES}

The basic experimental approaches used in the present experiments for
measurements of total and differential cross sections of ionization,
charge-exchange and excitation processes are the following: crossed-beam
spectroscopy method, refined version of a capacitor method, collision and
optical spectroscopy methods. The basic experimental set-up for measurements
of the total and differential cross sections of ionization, charge-exchange
and excitation processes in collisions between Na$^{+}$ ion and \ Ne and Ar
atoms was discussed previously in detail in Ref. \cite{Kez25}, so the only
brief description will be given here. The uniqueness of our experimental
approach \cite{Kez25} is that for colliding pairs Na$^{+}-$ Ne and Na$^{+}-$
Ar involved in the above mentioned processes the quality of the beam as well
as the experimental conditions always remain identical because the
experimental apparatus has three collision chambers. The latter allows to
use a refined version of the capacitor method, collision spectroscopy
method, and optical spectroscopy method under the same \textquotedblleft
umbrella.\textquotedblright\ For the experimental measurements of the
ionization cross section in collisions between Ne$^{+}$ and Ar$^{+}$ ions
and an alkali--metal atom a new experimental set-up is developed using a
crossed--beam spectroscopy method. Details of the apparatus and description
of the method are presented below.

\subsection{ Crossed--beam spectroscopy method}

\begin{figure}[t]
\begin{center}
\includegraphics[width=9cm]{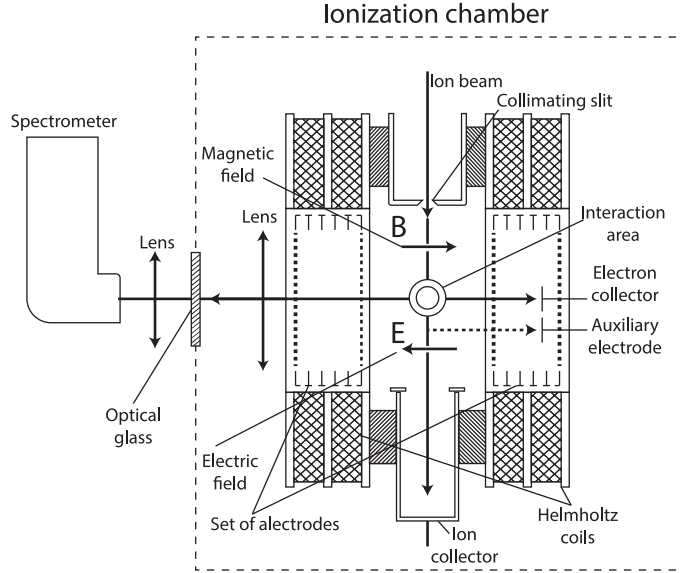}
\end{center}
\caption{ Schematic diagram of the experimental set-up for study of
collisions of ions with alkali metal atoms.}
\label{Fig1}
\end{figure}

Measurements of the ionization in collision of ions with alkali--metal atoms
are related to well known difficulties: preparation of the target;
determination of its density; protection of the surface of an ionization
chamber and insulators from desorption of a metallic vapor, etc. In the
present study, for measurements of the ionization cross section of alkali
metal atoms, we are using the method of intersected beams. This method of
measurement of the ionization cross section has some advantage compared to
other methods. Particularly, in the framework of the method, there is no
need to avoid the scattering of an incident beam (as it is peculiar e.g. for
a capacitor method) and secondary particles (recoil ions and electrons) on a
collector, the emission of electrons from the surface of collectors, etc.
The idea of crossed--beam method was suggested in Ref. \cite{Fite34}, and
used to study the ionization and charge transfer in proton--hydrogen atom
collisions. Here this method is significantly elaborated and for the first
time is used to investigate the ionization of alkali metal atoms. A
schematic drawing of the apparatus for the measurement of the ionization
cross section is shown in Fig. \ref{Fig1}. The core part of the experimental
set-up consists of the ion source, magnetic mass-analyzer, ionization
chamber, source of an atomic beam, system for collection of light from the
crossing beam region and spectrometer for analyzing of the radiation.

A primary ion beam from a 30 MHz radiofrequency ion source, using permanent
longitudinal magnetic field, passes through the single channel capillary
formed by the collimating slit and enters into the ionization chamber and is
collected by an ion collector. An input of high-frequency power in discharge
was carried out by an inductive connection. The power needed to cause the
discharge in the ion source was about 150 W. The formation of a plasma pinch
in the extraction region is possible by the Pyrex cup and stainless steel
capillary. The extraction voltage is 1.0-2.5 keV and the density of the ion
beam in the ionization chamber (after passing the magnetic mass--analyzer
and collimating slit) is about 0.1 mA/cm$^{2}$. A neutral atom beam emerged
from an aperture and passed through a collimating slit \ and directed into
the interaction area of the ionization chamber. The atomic beam itself is
formed by an extraction nozzle of an oven, where an alkali metal vapor is
generated. A diameter of the nozzle is 0.6 mm, while the length of the
channel is 6 mm. The oven is made from a tantalum with a massive cone which
is screwed on a crucible with the evaporated metal. The heating of the oven
is performed by the current of 1.8--2.0 A. The evaporated metal temperature
is measured and controlled by the thermocouple that is mounted on the oven.
At the exit from the oven at the distance of 10 mm from the nozzle the
density of the sodium atoms beam compose 0.3 $\times $10$^{11}$--3 $\times $%
10$^{11}$ cm$^{-3}$. The primary ion beam and the modulated atomic beam
cross each other perpendicularly at the interaction area of the ionization
chamber that is surrounded by magnetic and electric fields. The magnetic
field is created by the Helmholtz coils, while the uniform electric field is
produced by the set of circular electrodes. Both fields are parallel to each
other, opposite directed and perpendicular to the ion and the modulated
atomic beams. The atomic beam that is not shown in Fig. \ref{Fig1} is
perpendicular to the drawing plane. The magnetic field allows determination
of the crossing beam region from where the electrons are collected onto the
collector, while the uniform electric field is needed for transporting these
electrons towards the electron collector for the registration. Electrons
formed in the ionization chamber, as a result of collision of the ionic and
atomic beams are collected by the collector of electrons. The electrons,
those are produced mostly in the crossing region of the ion and atomic
beams, due to the presence of transverse magnetic field can escape from the
crossing region just transversely with respect to the direction of the ion
beam. The strength of the magnetic field accounts 100-200 Oersted. This
means that the shift towards the longitudinal direction does not exceed $1-2$
mm of the Larmor radius, and hence ensures to determine sufficiently
reliable the effective length of collection of electrons. The strength of
the electric field was chosen to maintain collection of all electrons with
the energy up to 23 eV. The last allows to detect electrons from
autoionization processes resulting in a liberation of the electrons with the
energy of about 20 eV that may be the dominant process in this collision.
However, let us mention, that the method allows the collection of electrons
with energy up to 45 eV, with the probability close to $90\%$. A current of
electrons on the collector is induced by the electrons produced due to the
ionization of the atomic beam and electrons of the \textquotedblleft
background\textquotedblright\ induced by the ion impact. For allot of a
partial current due to the ionization of \textquotedblleft background
\textquotedblleft atoms, it is possible to carry out a modulation of the
atomic beam or to use an auxiliary electrode, implemented in our apparatus.
The auxiliary electrode is shifted with respect to the main collector of
electrons, therefore for a correct orientation of the atomic beam the
detector can collect the electrons formed during interaction of ion beam
with the \textquotedblleft background\textquotedblright\ atoms.

For the measurements of the absolute ionization cross section it is needed
to know a geometric dimension of the interaction area, an effective length
of collected electrons produced in the collision, as well as it is necessary
to control a single collision condition, determine the density of particles
in the target beam and the flow of incident particles as well. For this
reason an optical channel is used that is located on the left side of the
ionization chamber. This channel consists of a lens, optical glass and
spectrometer with a photoelectron adaptor and photomultiplier that is
cooling by liquid nitrogen. The radiation from the beams' crossing region
that is produced due to the excitation of colliding particles is extracted
and collected perpendicularly to the beams along the axis of Helmholtz coil
on a side opposite to the collector of the electrons.

One of the problems, arising during the measurement of the absolute
ionization cross section of alkali--metal atoms, as was mentioned above, is
related to the determination of the density of the target beam. The method
of determination of the sodium atom beam density used in our work is based
on the measurement of the intensity of resonant lines of a Na atom excited
by a proton impact. We choose this approach for determination of target
density because the excitation cross section of the resonance line induced
by protons collision belongs to processes which cross sections are measured
more reliable \cite{Lavrov35, Shingel, Igenbergs37}. A special attention was
paid to the reliable determination and monitoring of the absolute spectral
sensitivity of the light recording system. This was done by a registration
of a Ne atom line ($\lambda =585.2$ nm, transition 3p [1/2]$_{0}$ -- 3s[1/2]$%
_{1}$) excited by the proton at an energy of 10 keV. The cross section of
this line is known \cite{Heer38}, and the wave length of the line
sufficiently close to the wave length of resonance lines of a Na doublet. As
to the concentration of Ne atoms in the region of collection of the
electrons, it could be measured by a simple and reliable method -- by
measuring the pressure of neon in the collision chamber, using an ionization
manometric lamp, calibrated by a compression manometer of Macleod gauge.

The sources of measurement uncertainties of the method are mostly related to
the nature of the vapor--metallic target, impossibility of quick shutdown of
the target, the increase of surface conductivity of an insulator due to
condensation of Na vapor, and the presence of heated elements in the
ionization chamber. All these uncertainties were minimized in our
measurements. The resultant uncertainty of the measurements is estimated as
15\%, and is linked mainly with the measurement of the Ne pressure, the
accuracy of which is estimated as 7\%.

\subsection{ Refined version of a capacitor method}

A beam of Na$^{+}$ ions from a surface-ionization ion source is accelerated
and focused by an ion -- optics system, which includes quadruple lenses and
collimated slits \cite{Kez25}. After the beam passes through a magnetic mass
spectrometer, it enters the collision chamber containing Ne and Ar gases.
The pressure in the collision chamber when there is no a Ne or Ar target gas
is kept at about 10$^{-6}$ Torr, while the typical pressure under operation
is 10$^{-4}$ Torr, which is low enough to ensure single--collision
conditions. The charge--exchange and ionization cross sections were measured
by a refined version of the capacitor method \cite{Kikiani26}. In an earlier
paper \cite{Ogurtsov4} the measurements were performed by the standard
transfer electric field method. It is the customary procedure to use one of
the central electrodes as the measurements electrode. We consider that such
an approach is the reason for significant errors in measurements \cite%
{Ogurtsov4} because scattered primary ions may strike the electrodes used
for measurements. To avoid this deficiency we used a refined version of the
transfer electric field method by shifting from the central electrode (a
standard method) to the first electrode (towards the beam entrance side). In
this case the defeat of the electrodes by the scattered primary ions that
affects the results of measurements is substantially reduced. Due to
fringing effects at the edges of this electrode a system of auxiliary
electrodes between the first electrode and the entrance slit were installed.
These auxiliary electrodes create a uniform potential near the first
electrode. The first electrode, the auxiliary electrodes, and the entrance
slit are all positioned together as close as possible. This close
arrangement limits the scattering region of the beam to the entrance side.
The primary ions are detected by the Faraday cup. The particles (secondary
positive ions and free electrons) produced during collision are detected by
a collector. The collector consists of two rows of plate electrodes that run
parallel to the primary ion beam. A uniform transverse electric field,
responsible for the extraction and collection of secondary particles, is
created by the potentials applied to the grids. This method yields direct
measurements of the cross section $\sigma ^{+}$ for the production of singly
positively charged ions and $\sigma ^{-}$ for electrons as the primary beam
passes through the gas under study. These measured quantities are related in
an obvious way to the capture cross section $\sigma _{c}$ and the apparent
ionization cross section $\sigma _{i}$ and are determined as

\begin{equation}
\sigma ^{+}=\sigma _{c}+\sigma _{i},\ \ \ \ \ \ \ \ \ \ \sigma ^{-}=\sigma
_{s}+\sigma _{i}.  \label{IonS}
\end{equation}%
In (\ref{IonS}) $\sigma _{s}$ is the stripping cross section of the incident
ion. The ionization cross section $\sigma _{i}$ is always larger than the
cross section for stripping $\sigma _{s}.$

For Na$^{+}-$ Ne collisions, the uncertainty in the measurements of
charge--exchange and ionization cross sections are estimated to be 15\%.
This is determined primarily by the uncertainty in the measurement of the
absolute values of the cross sections $\sigma ^{+}$ and $\sigma ^{-}$ and by
the uncertainty in the measurement of a target gas pressure in the collision
chamber.

For Na$^{+}-$Ar collisions, the uncertainty in the measurements of the
absolute values of the cross sections $\sigma ^{+}$ and $\sigma ^{-}$ is
estimated to be 15\% over the entire energy interval studied. This is
determined primarily by the uncertainty in the measurement of the target gas
pressure in the collision chamber. The uncertainty in the determination of
the ionization cross section $\sigma _{i}$, is estimated to be 15\% over the
energy range and it is determined by the error in the measurement of $\sigma
^{-}$. The uncertainty in the measurements of the capture cross sections $%
\sigma _{c}$ at the energy less than 2 keV is estimated to be 15\%, while at
the energy 5 keV and above the uncertainty does not exceed 25\%. For Na$%
^{+}- $ Ar collisions at the energy less than 2 keV the cross section $%
\sigma ^{+}$ is significantly larger than $\sigma ^{-}$. Accordingly, the
error in the determination of the capture cross section $\sigma _{c}$ in
this energy region is related primarily by the error in the measurement of $%
\sigma ^{+}.$ With increasing of the Na$^{+}$ beam energy the cross sections 
$\sigma ^{+}$ and $\sigma ^{-}$ becomes more nearly equal. As a result the
error in the determination of $\sigma _{c}$ increases.

\subsection{Collision spectroscopy method}

The energy--loss spectra and differential scattering experiments have been
performed with a collision spectroscopy apparatus. Since the details of the
apparatus have been given elsewhere \cite{Gochitashvili27}, only a brief
description will be given below.

The primary beam extracted from the ion source was accelerated to the
desired energy before being analyzed according to $q$/$m$ ($q$ and $m$ are
the ion's charge and mass, respectively). The analyzed ion beam was then
allowed to pass through the collision chamber by appropriately adjusting the
slits prior to entering into a \textquotedblleft box\textquotedblright\ type
electrostatic analyzer. The energy resolution of this analyzer is $\Delta
E/E $ = 1/500. Automatic adjustments of the analyzer potentials gives the
possibility for investigation of the energy--loss spectra in the energy
range of 0--100 eV. The differential cross section is measured by rotating
the analyzer around the center of collisions over an angular range between 0$%
^{\circ }$ and 25$^{\circ }$. The laboratory angle is determined with
respect to the primary ion beam axis with an accuracy of 0.2$^{\circ }$.

For the measurements of the charge--exchange differential cross section the
charge component of scattered primary particles realized in the collision
chamber is separated by the electric field and neutral particles formed by
electron--capture collisions are registered by the secondary electron
multiplier. Such a tool gives us the possibility to determine the total
cross sections and to compare them with the results obtained by the refined
version of the capacitor method \cite{Kikiani26}. In addition, the measured
energy--loss spectrum gives detailed information related to the intensity of
inelastic processes realized in the excitation, charge--exchange and
ionization processes.

\subsection{Optical spectroscopy method}

The method used for the optical measurements have been described previously 
\cite{Gochitashvili28}, therefore a brief description will be given here.
The alkali metal ion beam leaving the surface--ionization ion source is
first accelerated to a predetermined energy. It is then focused by the
quadruple lenses and analyzed by the mass spectrometer. The emerging ions
passed through a differentially pumped collision chamber containing the
target gas at low pressure. The ion current is measured by the collector and
the light emitted, as a result of the excitation of colliding particles,
from the collision chamber is viewed perpendicularly to the beam by a
spectrometer. The spectral analysis of the radiation was performed in the
vacuum ultraviolet as well as in the visible spectral regions. The linear
polarization of the emission in the visible part of the spectrum is analyzed
by the Polaroid and the mica quarter-wave phase plate in front of the
entrance slit of the monochromator. The phase plate is placed after the
polarizer, is rigidly coupled to it, and used to cancel the polarizing
effect of the monochromator. A photomultiplier tube with the cooled cathode
is used to analyze and detect the emitted light. The spectroscopic analysis
of the emission in the vacuum ultraviolet region is performed with the
Seya-Namioka vacuum monochromator, incorporating a toroidal diffraction
grating. The radiation was recorded by the secondary electron multiplier
used under integrating or pulse-counting conditions. The outputs of the
photomultiplier and the secondary electron multiplier were recorded by the
electrometers. The polarization of the radiation in the vacuum ultraviolet
was not taken into account. The absolute excitation cross sections for the
resonance lines of sodium that are determined by comparing the measured
output signal with one that due to the excitation of a nitrogen by an
electron impact. A particular attention is devoted to the reliable
determination and control of the relative and absolute spectral sensitivity
of the light--recording system. This was done by measuring the signal due to
the emission of molecular bands and atomic lines excited by electrons in
collisions with H$_{2}$, N$_{2}$, O$_{2}$, and Ar. For this, an electron gun
was placed directly in front of the entrance slit of the collision chamber.
The relative spectral sensitivity, and the values of the absolute cross
sections, is obtained by comparing the cross sections for the same lines and
molecular bands reported in Refs. \cite{Ajello29, Avakyan30, Stone31, Tan32,
Tan33}. The uncertainties in the excitation cross sections for the Na$^{+}-$
Ar system are estimated to be 20\% and the uncertainty of the relative
measurements does not exceed 5\%.

\section{EXPERIMENTAL RESULTS AND ANALYSIS}

\begin{figure}[t]
\begin{center}
\includegraphics[width=10cm]{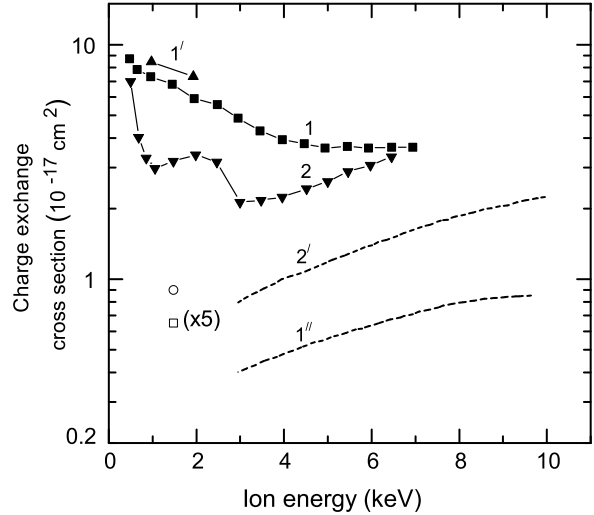}
\end{center}
\caption{ Dependences of the absolute charge--exchange cross sections on
energy of a Na$^{+}$ ion in Na$^{+}-$Ne and Na$^{+}$--Ar collisions. Curves:
1 -- Na$^{+}$--Ne, present data; 1$^{\prime }$ -- Na$^{+}-$Ne, data from
Ref. \protect\cite{Os22}; 1$^{\prime \prime }$ -- Na$^{+}-$Ne, data from
Ref. \protect\cite{Ogurtsov4}; 2 -- Na$^{+}-$Ar, present data; 2$^{\prime }$
-- Na$^{+}-$Ar, data from Ref. \protect\cite{Ogurtsov4}; $\circ $ -- Na$%
^{+}- $Ar, electron capture in resonance state, data from Ref. \protect\cite%
{Kita17}; $\square $ -- Na$^{+}-$Ar, electron capture with the excitation of
target ion, data from Ref. \protect\cite{Kita17} are multiplied by factor of
5. }
\label{Fig2}
\end{figure}

\begin{figure}[t]
\begin{center}
\includegraphics[width=10cm]{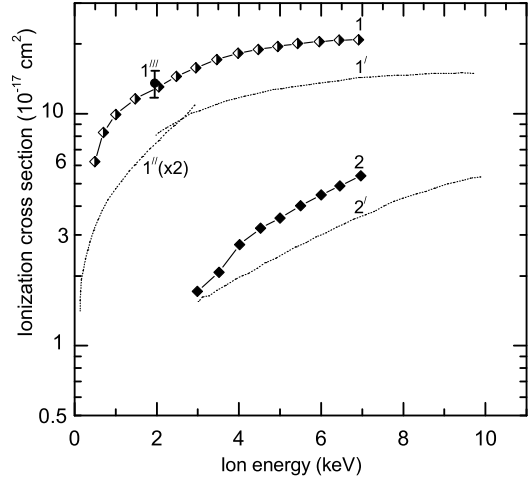}
\end{center}
\caption{ Dependences of the absolute ionization cross sections on energy of
Na$^{+}$ ion in Na$^{+}-$Ne and Na$^{+}-$Ar collisions. Curves: 1 -- Na$%
^{+}- $Ne, present data; 1$^{^{\prime }}$ -- Na$^{+}-$Ne, data from Ref. 
\protect\cite{Flaks1}; 1$^{^{\prime \prime }}$ -- Na$^{+}-$Ne, data from
Ref. \protect\cite{Latypov11} are multiplied by a factor of 2; 1$^{\prime
\prime \prime }$ -- Na$^{+}-$Ne, data from Ref. \protect\cite{Os22}, at
fixed $E=2$ keV; 2 -- Na$^{+}-$Ar, present data; 2$^{^{\prime }}$ -- Na$^{+}$%
--Ar, data from Ref. \protect\cite{Ogurtsov4}.}
\label{Fig3}
\end{figure}

\begin{figure}[t]
\begin{center}
\includegraphics[width=10cm]{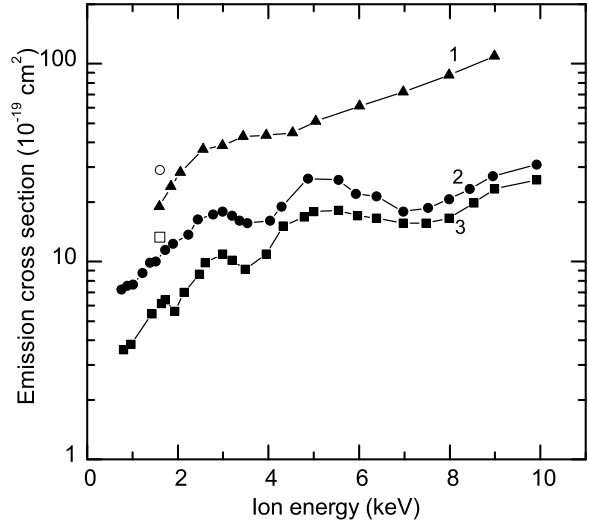}
\end{center}
\caption{ Excitation function for sodium and argon atomic lines in Na$^{+}-$%
Ar collisions. Curves: 1 -- NaI ($\protect\lambda $\ $=389.0-389.6$ nm,
3p--3s transition); 2 -- ArI ($\protect\lambda =104.8$ nm, 4s$^{^{\prime }}$%
--3p transition); 3 -- ArI ($\protect\lambda =106.7$ nm, 4s--3p transition); 
$\circ $ and $\square $ denote the excitation of sodium and argon atoms,
respectively, obtained in Ref. \protect\cite{Kita17}. }
\label{Fig4}
\end{figure}

\begin{figure}[t]
\begin{center}
\includegraphics[width=10cm]{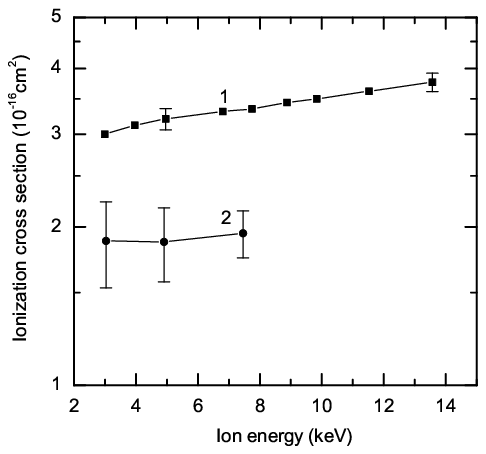}
\end{center}
\caption{ Dependence of the absolute ionization cross sections on energy of
Ne$^{+}$ and Ar$^{+}$ ions in Ne$^{+}-$Na and Ar$^{+}-$Na collisions.
Curves: 1 -- Ne$^{+}-$Na; 2 -- Ar$^{+}-$Na.}
\label{Fig5}
\end{figure}

\begin{figure}[t]
\begin{center}
\includegraphics[width=10cm]{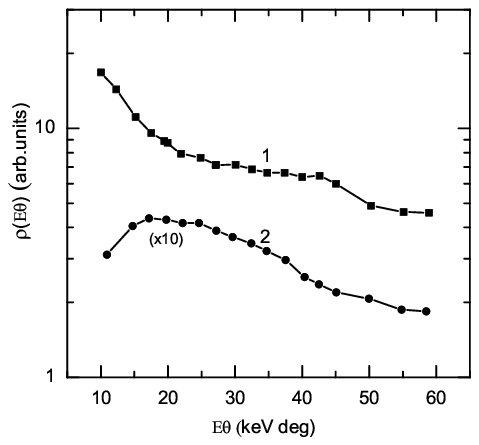} 
\end{center}
\caption{Differential cross sections for Na$^{+}-$Ar collisions as a
function of reduced scattering angles at a beam energy $E=5$ keV. $%
\blacksquare $ -- the DCS of the elastic scattering. $\bullet $ --
excitation of the Rydberg states of Ar [3p$^{5}$ ($^{2}$P)4p; 3p$^{5}$($^{2}$%
P)3d]. The later data are multiplied by factor of 10. }
\label{Fig6}
\end{figure}

In what follows, the results for Na$^{+}$ion collisions with the Ne and Ar
atoms and Ne$^{+}$ and Ar$^{+}$ ions collisions with the alkali metal target
atom Na are presented and the findings are compared with data from the
literature. The results for the cross section measurements are shown in
Figs. 2--6. The energy dependences of the charge--exchange, ionization and
excitation for Na$^{+}$--Ne and Na$^{+}$--Ar collisions are presented in
Figs. \ref{Fig2}, \ref{Fig3} and \ref{Fig4}, respectively. Here for the
comparison we also present data of other authors. Results of the first
measurement of the energy dependence of ionization cross sections for the
processes (\ref{NeNa2}) and (\ref{ArNa2}) are presented in Fig. \ref{Fig5}.
Fig. \ref{Fig6}, where we plot the reduced cross section $\varrho =$ $\theta
\sin \theta \sigma (\theta )$ versus reduced angle $\tau =$ $E\theta ,$
shows the typical example of angular and energy dependences of differential
cross sections in the laboratory system when Na$^{+}$ ions are scattering on
the Ar atoms at a fixed energy $E=5$ keV$.$ The reduced scattering angle is
defined as $\tau =E\theta ,$ where $E$ is the energy of the incident beam in
keV and $\theta $ is a scattering angle in degrees. The filled squires in
Fig. \ref{Fig6} represent the elastic scattering of the Na$^{+}$ ions, while
by the solid circles are shown results of the direct excitation of the Ar
atoms in 4p and 3d Rydberg states. In addition, we estimated the electron
energies released in Na$^{+}-$Ne and Na$^{+}-$Ar collisions. The estimates
were obtained from the measurements of the dependences of electron current
in the measuring electrodes on the potential applied to these electrodes for
the collection of electrons. It was found that the energy of the most
liberated electrons is below $10-15$ eV.

The results for the charge-exchange cross section for Na$^{+}-$Ne and Na$%
^{+}-$Ar collisions along with the data from literature are shown in Fig. %
\ref{Fig2}. The comparison of our measurements for the charge--exchange
cross section with the results obtained in \cite{Os22} at two fixed energy $%
E=1$ keV and $E=2$ keV shows excellent agreement. However, a dramatic
difference by about 2 orders in the magnitude, as well as in the behavior of
the energy dependences are observed, when one compares our results with the
results obtained in \cite{Ogurtsov4}. The same tendency, but the discrepancy
in the magnitude by about 1 order observed when one compares our results for
the Na$^{+}-$Ar pair with the data from Ref. \cite{Ogurtsov4}. Our results
for the charge--exchange processes (\ref{Ne1}), (\ref{Ar1}) can be compared
with the cross sections obtained in Ref. \cite{Kita17} by integrating the
differential cross section over the scattering angle for an electron capture
and a capture with the excitation of a target ion at energy of $E=1.5$ keV,
which are $9.0\times 10^{-18}$ cm$^{2}$ and $1.3\times 10^{-18}$ cm$^{2},$
respectively. This comparison shows that the discrepancy is threefold.

The results for the ionization cross section for Na$^{+}-$Ne and Na$^{+}-$Ar
collisions along with the data of previous measurements are shown in Fig. %
\ref{Fig3}. The comparison of our ionization cross section for the process (%
\ref{Ar2}) with the results obtained in \cite{Ogurtsov4} shows a
satisfactory agreement at low energies but the discrepancy increases for the
energies of ions $E>4$ keV. In principle, a satisfactory agreement is
observed for a Na$^{+}-$Ne collision between our results and the results
from \cite{Flaks1}, especially in the energy dependence of the cross
sections. A rather significant discrepancy is observed if one compares our
results with the results reported in Ref. \cite{Latypov11}. Within the
accuracy of measurements an excellent agreement is observed between our
results and the results obtained in Ref. \cite{Os22} at a fixed ion energy $%
E=2$ keV.

The excitation function for Na$^{+}-$Ar collisions is presented in Fig. \ref%
{Fig4}. Our data of the excitation function for the resonance lines of
sodium (curve 1) and argon atoms (curves 2 and 3) can be compared only with
the results obtained in Ref. \cite{Kita17} and only for the collision energy
of $E=1.5$ keV (open circle and open square in Fig. \ref{Fig4}). It can be
seen from Fig. \ref{Fig4} that the measured total excitation cross sections
of a sodium atom in 3p state (curve 1) and argon atom in 4s and 4s$%
^{^{\prime }}$ states (curves 2 and 3, respectively) are in reasonable
agreement with the data obtained in \cite{Kita17}.

To the best of the authors' knowledge in the energy range considered there
are no experimental measurements of the ionization cross sections for the
process of ionization of alkali metal Na by Ne$^{+}$ and Ar$^{+}$ ions. The
first measurements of the absolute ionization cross sections for Ne$^{+}-$Na
and Ar$^{+}-$Na collisions are presented in Fig. \ref{Fig5}. As it seen from
Fig. \ref{Fig5} the value of the ionization cross section strongly depends
on the mass ratio of the colliding particles. In case of Ne$^{+}-$Na
collisions when the mass ratio of the colliding particles is close to 1, the
cross section is larger by a factor of 2 in comparison to the asymmetric
case of Ar$^{+}-$Na collisions when the mass ratio is greater than 1.

Distinct features are observed in the differential cross section (DCS) as a
function of the reduce scattering angle $\tau $ shown in Fig. \ref{Fig6}.
The DCS with the excitation of Ar atoms is smaller than the DCS for the
elastic scattering. Another feature is the alternative behavior of the DCS
at small angles: the elastic scattering increases at the small angles, while
the DCS of the Rydberg states of Ar decreases. The most striking feature is
that the ratio of the elastic scattering to the excitation cross sections
strongly increases for small values of $\tau ,$ while for $\tau >20$ only
varies relatively weakly.

The data obtained in this study can be used to make certain conclusions
related to possible mechanisms of the investigated processes. To explain
these mechanisms one can use a schematic quasimolecular terms for the system
of colliding particles. The quasimolecular nature of the interaction of the
above considered collision partners can be visually manifested by a
representative Na$^{+}-$ Ar colliding pair. Therefore, below we discuss the
mechanisms of the realized processes for this colliding pair. The
corresponding schematic correlation diagram constructed based on Bara --
Lichten rules \cite{Barat} is presented in Fig. \ref{Fig7}.

\subsection{Charge exchange in Na$^{+}-$Ar collisions}

\begin{figure}[t]
\begin{center}
\includegraphics[width=10cm]{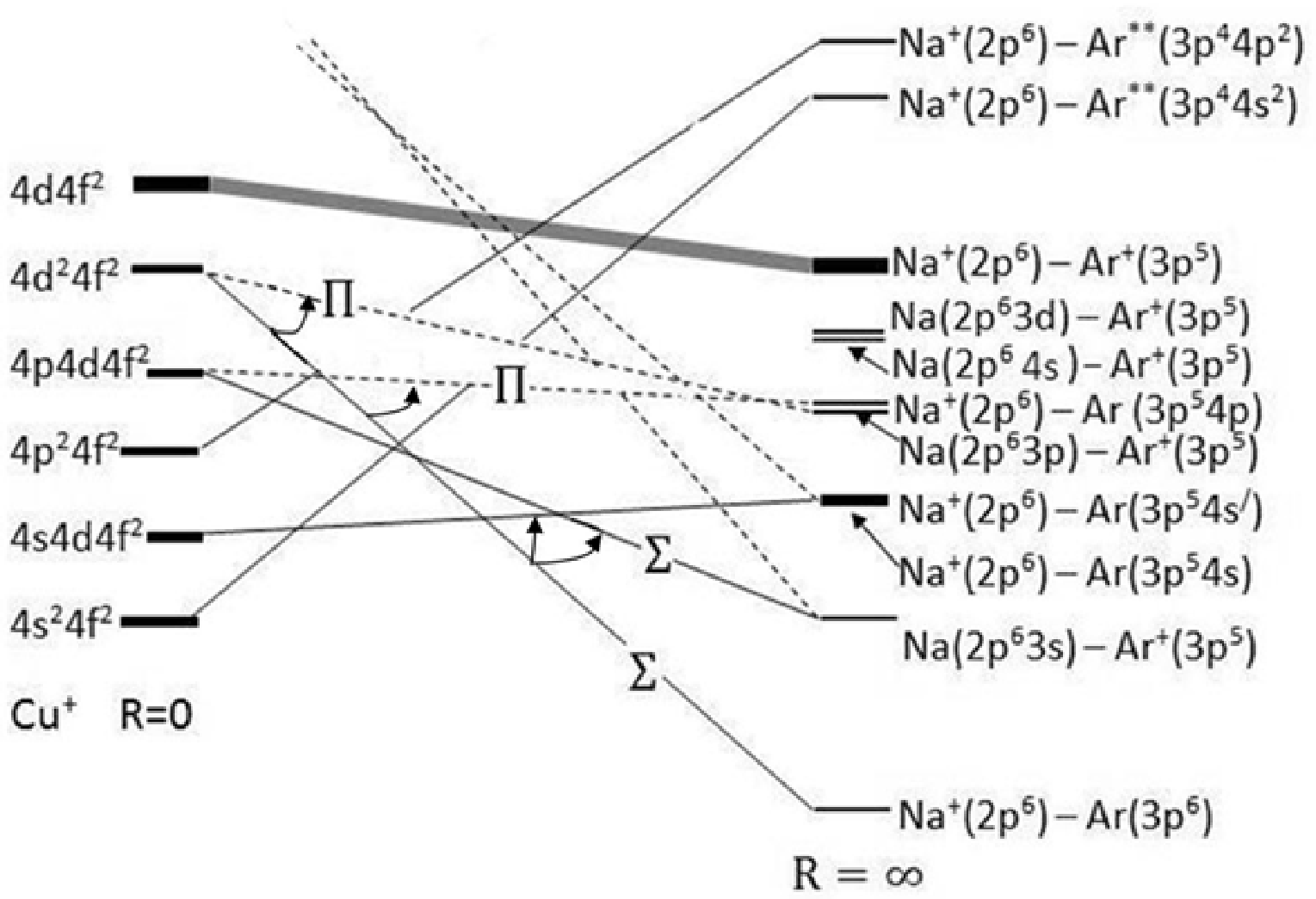} 
\end{center}
\caption{Schematic correlation diagram for Na$^{+}-$Ar colliding pair. Solid
lines indicate $\Sigma $ states, dashed lines indicate $\Pi $ state.}
\label{Fig7}
\end{figure}

For determining the processes responsible for the charge exchange in Na$%
^{+}- $Ar collisions, we compare the total charge exchange cross sections
(curve 2 in Fig \ref{Fig2}.) with the total cross section of radiation for
the \ resonant lines $\lambda =389.0-389.6$ nm of a Na atom (curve 1 in Fig. %
\ref{Fig4}). Taking into account the selection rules and the ratio of the
oscillator strengths for these transitions, one can prove that the emission
of radiation from any level of the sodium atom culminates in the half of
cases in transition of the atom to the resonant state, followed by emission.
Consequently, the doubled deexcitation cross section of a Na atom gives a
clue related to the capture cross section in the excited state. It can be
seen from Fig. \ref{Fig4}. (curve 1) that the emission cross section of
resonant levels of sodium atoms in Na$^{+}-$Ar collisions increases with the
increase of the ion energy, amounting to $\backsim $1.8$\times $10$^{-18}$ cm%
$^{2}$ for the ion energy $E=1.5$ keV and $\backsim $6$\times $10$^{-18}$ cm$%
^{2}$ for $E=5-7$ keV. A comparison of the behavior of the two curves leads
to the conclusion that two nonadiabatic regions responsible for an electron
capture: one at the energy range $E=0.5-2.0$ keV and the second one -- at
the energy range $E=3-7$ keV. In the collisions considered here, the
electron capture at low energies (up to $E=2$ keV) takes place as a result
of the electron capture into the ground state of a Na atom with the
formation of the argon ion in the ground state as well. The process realizes
through the channel Na$^{+}$(2p$^{6}$)$+$Ar(3p$^{6}$)$\rightarrow $Na(3s)$+$%
Ar$^{+}$ (3p$^{5}$), with the energy defect of $\Delta E=10.6$ eV
(one-electron process) and the electron capture into the ground state of the
sodium atom with the formation of the argon ion in the excited state, Na$%
^{+} $(2p$^{6}$)$+$Ar(3p$^{6}$)$\rightarrow $Na(3s)$+$Ar$^{+}$(3p$^{4}$ ($%
^{1}$D)4s]$-$29.1 eV (two-electron process) \cite{Kita17}. This process can
occur, as can be seen from the diagram in Fig. \ref{Fig7}, as a result of
the direct pseudo-crossing of the term corresponding to the state Na(3s)+Ar$%
^{+}$(3p$^{5}$) with the ground state of the system Na$^{+}$(2p$^{6}$)$+$%
Ar(3p$^{6}$). Since the Na$^{+}$(2p$^{6}$)$+$Ar(3p$^{6}$) state has only $%
\Sigma $ symmetry it follows that the $\Sigma -\Sigma $ transition play a
dominant role in the low--energy charge--exchange processes. At the energy
range $E>5$ keV, the electron capture into the excited state Na(3p) of the
sodium atom, with the formation of the argon atom in the ground state takes
place. This process that is attributed to the reaction Na$^{+}$(2p$^{6}$)$+$%
Ar(3p$^{6}$)$\rightarrow $Na(3p)$+$Ar$^{+}$(3p$^{5}$) $-$ $12.7$ eV plays
the dominating role. The $\Sigma -\Pi $ transitions (see Fig. \ref{Fig7})
associated with the rotation of internuclear axis play a certain role in the
electron capture to the excited 3p states.

\subsection{Ionization in Na$^{+}$--Ar collisions}

Experiments show that the mechanism of ionization in Na$^{+}$--Ar collisions
is characterized by the release of predominantly slow electrons with the
energies $E<15$ eV. In order to determine the channel and mechanism of
ionization, we estimate the contribution of several inelastic processes that
result in emission of slow electrons. To estimate the contribution of direct
ionization, we calculated the cross section of this process using the
results obtained in \cite{Solov'ev40}. According to \cite{Solov'ev40} in the
limit of an united atom, the process of ionization is associated with the
emergence of the diabatic energy level to the continuum in the range of the
nonadiabatic interaction of molecular orbitals with the same orbital angular
momentum. Analysis of correlations of molecular orbitals in the Na$^{+}$--Ar
system shows (Fig. \ref{Fig7}) that the 3p electrons of Ar atom, whose
ionization is considered, in the limit of the united atom correspond to the
4d electrons of the Cu$^{+}$ ion. Thus, for the estimated cross section we
choose the orbital angular momentum $l=2$. The binding energy $E_{nl}$ of
electrons in the nonadiabaticity region was assumed to be equal to the
binding energy of 4d electrons of the Cu$^{+}$ ion. The effective charge $%
Z_{eff}$ was determined by the interpolation of the results obtained by
Hartree in Ref. \cite{Hartree41}. For the 4d electrons of the Cu$^{+}$, we
obtained $Z_{eff}$ $=3.1$. The calculation of the direct ionization cross
section with these parameters proves that the contribution of this process
is less than 1\% for the energy of 3 keV of the sodium ions and does not
exceed $10\%$ for the energy of 6 keV. The same approach is applied to
determine the contribution of electrons yield as a result of stripping of
the projectile ions to the measured ionization cross section. As it can be
seen from the correlation diagram for \ the energy level in Fig. \ref{Fig7},
the 2p electrons of the Na$^{+}$ ion correlate with the 4f electrons of the
Cu$^{+}$ ion. For this reason, to estimate the cross section we chose the
value for the orbital angular momentum $l=3$ and assumed that $Z_{eff}$ =
3.1. In other words, the same as for the 4d electrons of the Cu$^{+}$ ion.
One also should estimate the contribution of the stripping process. As a
result of calculation we found that the contribution of stripping to the
total electron yield cross sections is $0.1\%$ for an ion energy of 3 keV
and less than $4.5\%$ for the ion energy of 6 keV. Consequently, we can
conclude that the contribution of these processes to the ionization cross
section is insignificant in the entire energy range.

The double ionization of Ar atom and capture accompanied by ionization of Ar
ion evidently make a small contribution to the ionization cross section.
There are two reasons for this: the absence of pseudo-crossings of the
corresponding quasimolecular terms with the ground state term as it seen
from diagram in Fig. \ref{Fig7}, and the large energy defect for these
processes, $43.4$ eV and $38.3$ eV, respectively.

By comparing consecutively other possible mechanisms of the release of
electrons with energy less than $10-15$ eV, we draw the conclusion that the
main mechanism of their emergence (along with the contribution of other
channels) is associated with the decay of autoionization states in an
isolated atom. According to Refs. \cite{Jrgensen20} and \cite{Kita17}, these
states are those with two excited electrons Na$^{+}$(2p$^{6}$)$+$Ar(4s)$%
\rightarrow $Na$^{+}$(2p$^{6}$)$+$Ar[ 3p$^{4}$(1D)4s$^{2}$; 4p$^{2}$] $-28.7$
eV .

\subsection{Excitation processes in Na$^{+}-$Ar collisions}

Let us consider the excitation mechanisms in Na$^{+}-$Ar collisions. The
most interesting is exploration of the excitation function in Fig. \ref{Fig4}
for the argon atom lines $\lambda =104.8$ nm and $\lambda =106.7$ nm for the
transitions 4s, 4s$^{^{\prime }}$--3p that show oscillatory structures
(curve 2 and 3, respectively). It can be seen from the correlation diagram
in Fig. \ref{Fig7}, the excitation of the Ar(4s) state can occur as a result
of i) $\Sigma -\Sigma $ transition between the entrance energy level [Na$%
^{+} $(2p$^{6}$)$-$Ar (3p$^{6}$)] and the level corresponding to the
excitation of the argon atom [Na$^{+}$(2p$^{6}$)$-$Ar (3p$^{5}$4s, 4s$^{%
{\acute{}}%
}$)] or ii) due to the 4p--4s cascade transition in an isolated atom. In the
later case the state 4p of the excited argon atom is due to the rotational $%
\Sigma -\Pi $ transition at small internuclear distances. Indeed, the
observed oscillatory structure, a comparatively small cross section $\sigma
\thicksim 10^{-18}$ cm$^{2}$, and a large oscillation depth of the curves 2
and 3 in Fig. \ref{Fig4} indicate, according to \cite{Bobashev42}, that
namely the contribution of the rotational $\Sigma -\Pi $ transition, with
population of the 4p energy level of Ar atom, should be significant in the
excitation process. Our results for the differential cross sections
presented in Fig. \ref{Fig6} just are the evidence of this fact. Moreover,
we have to mention, that among the various channels studied, just the
elastic scattering channel and excitation of the Rydberg states 4p and 3d of
the argon atom are populated effectively. As to the oscillatory behavior of
the data 2 and 3 in Fig. \ref{Fig4}, these oscillations are due to the
interference of very close quasimolecular states of the Na$^{+}$(2p$^{5}$)$-$%
Ar (3p$^{5}$4p) and Na(2p$^{6}$3p)$-$Ar$^{+}$(3p$^{5}$) systems that have
the energy scale defect only $0.19$ eV. But, if it so, and that is a rule,
in accordance with the interference model \cite{Ankudinov43}, the energy
dependence of the excitation cross section for the Na(2p$^{6}$3p) presented
by data 1 in Fig. \ref{Fig4} under the such assumptions may exhibit an
oscillatory structure in antiphase that gives the dependence observed for
the 4s and 4s$^{^{\prime }}$ lines of the Ar atom. However, the energy
dependence of the excitation cross section of the Na(2p$^{6}$3p) atom for
the 3p--3s transition with $\lambda =389.0-389.6$ nm in Fig. \ref{Fig4}
exhibits only a small structural singularity. This leads to the conclusion,
that other channels are contributed to the excitation of the Na(2p$^{6}$3p)
state, which can smoothen the oscillatory structure. Therefore, special
measures should be taken to explore this irrelevance. It was found, that the
flatness of the excitation cross section may occur due to the influence of a
cascade transition from the upper sodium levels (e.g. from 4s$^{2}$S$_{1/2}$
and 3d$^{2}$D$_{3/2}$) to the sodium 3p level. This assumption was verified
indirectly by us from analysis of the ratio of the excitation cross section
of the sodium singlet 3p$^{2}$P$_{1/2}$ and triplet 3p$^{3}$P$_{3/2}$ states.

It was revealed that the cross section ratio $\sigma $(3p$^{2}$P$_{1/2}$)$%
/\sigma $(3p$^{2}$P$_{3/2}$) differs from the statistical population in the
entire energy range and amounts to $\thicksim 0.7$ instead of $0.5$. The
probabilities of the cascade electron transitions from sodium 4s$^{2}$S$%
_{1/2}$ and 3d$^{2}$D$_{3/2}$ levels to the 3p level of sodium atom are such
that the transition from the 4s$^{2}$S$_{1/2}$ level to the singlet, as well
as to the triplet states, is the same and changes the statistical population
just insignificantly. However, the transition from the 3d$^{2}$D$_{3/2}$
level to the sodium singlet 3p level is five times higher compare to the
triplet level and, hence, increases the statistical population
significantly. Accordingly, from our estimation, and by taken into
consideration that the excitation function for the sodium 4s and 3p states
are the same (this is a relevant because of the defect $\backsim 0.4$ eV) we
can conclude that the absence of a clearly manifested oscillatory structure
on the excitation cross section of the Na atom lines can be associated with
the effect of the cascade transition from the upper levels.

As to the origin of oscillation, observed on an excitation function for the
Ar(4s, 4s$^{^{\prime }}$), to our mind it is caused not by the direct
excitation of the 4s and 4s$^{^{\prime }}$ states of the argon atom. The
excitation of argon atom takes place into the 4p state and than from here,
through the 4p$-$4s cascade transition it become apparent in the excitation
of the 4s and 4s$^{^{\prime }}$ states.

\section{DETERMINATION OF THE TRANSITION PROBABILITIES BETWEEN POTENTIAL
CURVES OF QUASIMOLECULAR SYSTEM}

\begin{figure}[t]
\begin{center}
\includegraphics[width=10cm]{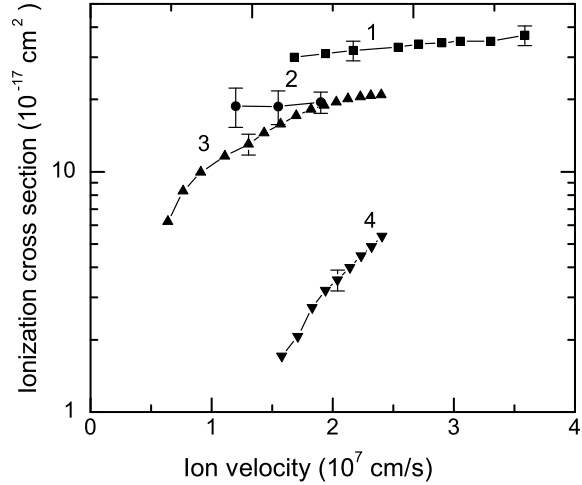} 
\end{center}
\caption{Velocity dependences of the ionization cross sections for the (NaNe)%
$^{+}$ and (NaAr)$^{+}$ systems. The result of the measurements of the
ionization cross section are presented for the following collisions: curve 1
-- Ne$^{+}-$Na; curve 2 -- Ar$^{+}-$Na; curve 3 -- Na$^{+}-$Ne; curve 4 -- Na%
$^{+}-$Ar.}
\label{Fig8}
\end{figure}

Let us compare the ionization processes (\ref{Ne2}), (\ref{Ar2}), (\ref%
{NeNa2}) and (\ref{ArNa2}). The velocity dependence of the ionization cross
sections for Na$^{+}-$Ne, Na$^{+}-$Ar, Ne$^{+}-$Na and Ar$^{+}-$Na
collisions are presented in Fig. \ref{Fig8}.\ As is seen from Fig. \ref{Fig8}
the magnitude of ionization cross sections strongly depends on the
ionization energy of the target atom. The ionization energy of Na, Ar and Ne
atoms are $4.8$ eV, $15.7$ eV, and $20.2$ eV, respectively. The lesser the
target atom's ionization energy, the bigger is the value of the cross
section. This fact is illustrated by the comparison of curves 1 and 3, and 2
and 4, in Fig. \ref{Fig8}.

The excitation processes for Na$^{+}-$Ne collisions are qualitatively
interpreted by the electron promotion model in Refs. \cite{Os22, Fano24}. In
order to discuss quantitatively the excitation mechanisms, one has to
evaluate the crossing parameters by collision experiments or \emph{ab initio}
calculations. Unfortunately, the accuracy of the calculations is still not
sufficient for many-electron systems. One of the objectives of this work is
to show that in some cases (e.g. for Na$^{+}-$Ne) the information extracted
from a theoretical calculation (e.g. the coupling matrix element) can be
obtained experimentally using a simple approach, namely by measuring the
ionization cross section of Na$^{+}-$Ne and Ne$^{+}-$Na colliding pairs in a
sufficient energy region.

\begin{figure}[t]
\begin{center}
\includegraphics[width=10cm]{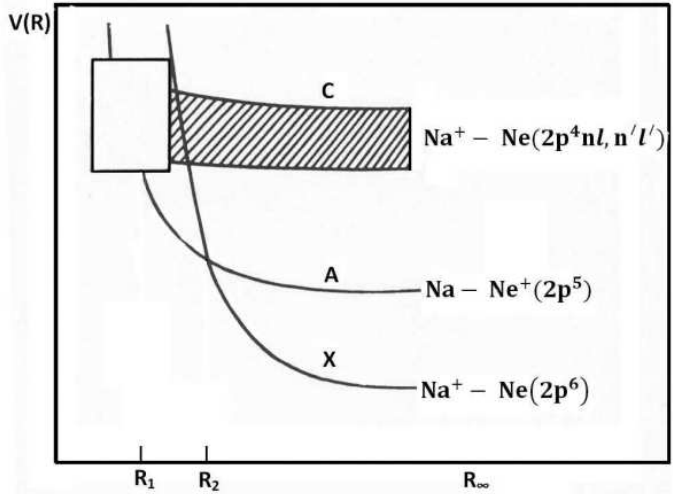} 
\end{center}
\caption{Schematic presentation of the potential terms for determination of
the transition probability and coupling matrix element for (NaNe)$^{+}$
system. "X" is the entrance potential curve. "A" denotes the potential curve
that corresponds to the charge exchange process. "C" represents a band of
terms corresponding to autoionization states.}
\label{Fig9}
\end{figure}

Usually for determination of transition probabilities for quasimolecular
systems in a region of pseudo--crossing of the potential curves one measures
the cross section of an inelastic transition between the states,
corresponding to these potential curves. What will be shown below, in some
cases this probability can be determined, not through a measurement of the
cross section for a transition from one channel to the other, but using two
independent measurements: the transition from one and, independently, from
other channels to a third channel. In this case, for determination of the
transition probability it is fully sufficient to determine not the absolute
value of cross sections, but only their relation.

Let us consider this method of determination of the transition probability
between potential curves, corresponding to the ground state (potential curve
X in Fig. \ref{Fig9}) and the states, in which particles are
charge-transferred (potential curve A in Fig. \ref{Fig9}) for the system
(NaNe)$^{+}$. As the third channel, in which the transition from these two
states occurs in considering collisions, can be chosen a channel of atomic
autoionization terms (band autoionization states C in Fig. \ref{Fig9}). In
accordance of choosing of the third channel, it is necessary to have the
cross sections for the ionization in Na$^{+}-$Ne and Ne$^{+}-$Na collisions
at the same collision velocity. Such data are obtained in this study. The
measurements of the ionization cross sections for the processes (\ref{Ne2})
and (\ref{NeNa2}) are brought specially to be realized the considered method.

In Ref. \cite{Os22} it was shown, that the ionization in collision of Na$%
^{+}-$Ne is realized as a result of the sequence transitions, at first due
to a pseudo--crossing of the potential curves X and A in the area of $R_{2}$
and then by the potential curve A with the band of curves C in the area $%
R_{1}$ as is shown in Fig. \ref{Fig9}. In this case, as it easy to see, the
mechanism of ionization in Na$^{+}-$Ne collision is the same, as in Ne$^{+}-$%
Na collision. The difference in the cross sections of ionization for these
pairs are related to the way the system approaches to the pseudo--crossing
region: in one case, by the potential curve X, while in the other case,
along the potential curve A. Take this fact into the consideration, in a
framework of an impact parameter approach, the cross section of ionization
in collisions of Na$^{+}-$Ne and Ne$^{+}-$Na pairs, at the same velocity of
relative motion, can be presented, as

\begin{equation}
\sigma _{1}=2\pi (1-P_{0})\int P_{AC}(b)bdb,  \label{NaNee}
\end{equation}%
\begin{equation}
\sigma _{2}=2\pi P_{0}\int P_{AC}(b)bdb,  \label{NeNae}
\end{equation}%
where $P_{AC}$ is the transition probability at the pseudo--crossing of the
potential curve A with the band C in the area of $R_{1}$ and $P_{0}$ is the
transition probability between the potential curves X and A in the area of $%
R_{2}$, at the some value of the impact parameter $b$ from the region $b\leq
R_{1}$. The later condition is clear because if this will be not satisfied,
particles never reach the region of $R_{1}$ and the ionization will not be
realized. The transition probability $P_{0}$ in the expressions (\ref{NaNee}%
) and (\ref{NeNae}) for the cross sections should be under the integral
because it is the function of the impact parameter -- $P_{0}(b)$. However,
since location of non-adiabatic area is such that $R_{1}<R_{2}$, the
dependence of $P_{0}(b)$ on the impact parameter $b$ for $b\leq R_{1}$ is
weak and therefore it is physically reasonable to consider $P_{0}(b)=P_{0}$
and pull out from the integral. Since the dependence of the probability $%
P_{0}$ on the impact parameter $b$ is linked to the radial velocity in the
transition region, it is possible to estimate a value of the radial velocity
sufficiently precisely, corresponding to these probabilities.

From the comparison of the ionization cross sections for Na$^{+}$--Ne and Ne$%
^{+}$--Na collisions is obtained that in collision of (NaNe)$^{+}$ at the
energy $7.0$ keV, probability $P_{0}=0.62$. To this value of $P_{0}$, for
the values of $R_{1}$ and $R_{2}$ from the article \cite{Os22}, corresponds
the radial velocity $V_{R}=0.7v_{0}$, where $v_{0}$ is the velocity of
relative motion of the colliding particles. Now by knowing the transition
probability and behavior of the potential curves in a non-adiabatic region
one can find the coupling matrix element $H_{XA}$ for nonadiabatic states
using the Landau-Zener formula \cite{Landau, Zener} for the probability of a
nonadiabatic transition for the pseudo--crossing potential curves

\begin{equation}
P_{0}=\exp \left( 2\pi \left\vert H_{XA}\right\vert ^{2}/V_{R}\left\vert
\Delta F\right\vert \right) .  \label{LZ}
\end{equation}%
In Eq. (\ref{LZ}) $\Delta F$ is the difference of slopes of the intersecting
potential curves. Taking the difference of the slopes $\Delta F=$3 a.u. from
Ref. \cite{Os22}, for the coupling matrix element $H_{XA}$ one gets $%
H_{XA}=0.14$ a.u. which significantly clarifies the theoretical estimation
of the value of this matrix element $H_{XA}=0.04-0.1$ a.u. obtained in Ref. 
\cite{Os22}. Despite its limitations, the Landau-Zener formula remains an
important tool for a nonadiabatic transition. Even in systems for which
accurate calculations are possible, application of the Landau-Zener formula
can provide useful estimates of nonadiabatic transition probabilities.
Alternatively, if the nonadiabatic transition probabilities and slopes are
known, this equation offer a feasible way to obtain the coupling matrix
element.

\section{SUMMARY AND CONCLUSIONS}

In this work, we report the results of the experimental study of inelastic
processes realized in collisions of Na$^{+}$ ion with Ne and Ar atoms and Ne$%
^{+}$ and Ar$^{+}$ ions with Na atoms in the impact energy range $0.5-14$
keV. In case of Na$^{+}$ ion Ne, Ar atoms collisions the absolute value of
the ionization, charge-exchange and excitation cross section are measured at
the energy range of $0.5-10$ keV, while in the case of Ne$^{+}$ and Ar$^{+}$
ion collision with Na the ionization cross section is measured at the energy
range of $2-14$ keV.

Using the experimental set-up based on the crossed-beam spectroscopy method
and the unique experimental set-up that includes a refined version of the
capacitor method, collision spectroscopy and optical spectroscopy methods of
measurements under the same umbrella, and a well--checked calibration
procedure of the light recording system, we have measured the absolute
values of the charge-exchange, ionization and excitation cross sections for
(NaNe)$^{+}$ and (NaAr)$^{+}$ systems. The correlation diagram of the (NaAr)$%
^{+}$ system has been employed to discuss the mechanism realized in Na$^{+}-$%
Ar collisions.

For the charge--exchange processes two nonadiabatic regions was revealed in
Na$^{+}-$Ar collisions. One region is at low energy, $E<2$ keV, where the
charge--exchange realizes as a result of electron capture into the ground
state of the sodium atom with the formation of the argon atom in the ground
state also in the region of pseudo-crossing of the potential curves of $%
^{1}\Sigma $ symmetry. While the other one is in the energy region $E>$3
keV, where the electron capture into the excited 3p state of a sodium atom
takes place and the formation of an argon ion in the ground state plays a
dominant role. In this case the $\Sigma -\Pi $ transition is realized and it
is associated with the rotation of the internuclear axis.

A primary ionization mechanism for Na$^{+}-$Ar colliding pair is related to
the liberation of slow electrons with the energy of $10-15$ eV and is
associated with the decay of autoionization states in an isolated atom.

The oscillatory behavior of the energy dependence of the excitation cross
section of argon atoms is revealed in Na$^{+}$--Ar collisions and found
that, this excitation is a result of the $\Sigma -\Pi $ transition between
the entrance energy level and the level corresponding to the excitation of
Ar atoms, and also due to the 4p--4s cascade transition in the isolated atom.

Experimentally measured ionization cross sections, for Na$^{+}-$Ne and Ne$%
^{+}-$Na colliding pairs in conjunction with the Landau-Zener formula, allow
us to determine the coupling matrix element and transition probability in a
region of pseudo-crossing of the potential curves.

\section*{Acknowledgements}

This work was supported by the Georgian National Science Foundation under
the Grant No.31/29 (Reference No. Fr/219/6-195/12). R.Ya.K. and R.A.L
gratefully acknowledge support from the International Reserch Travel Award
Program of the American Physical Society, USA.


\begin{thebibliography}{99}
\bibitem{Mouzon9} J. P. Mouzon, Phys. Rev. \textbf{41}, 605 (1932).

\bibitem{Moe5} D. E. Moe and O. H. Petsch, Phys. Rev. \textbf{110}, 1358
(1958).

\bibitem{Flaks1} I. P. Flaks, B. I. Kikiani, and G. N. Ogurtsov, Tech. Phys. 
\textbf{10}, 1590 (1966).

\bibitem{Ogurtsov4} G. N. Ogurtsov and B. I. Kikiani, Tech. Phys. \textbf{11}%
, 362 (1966).

\bibitem{Matveev7} V. B. Matveev, S. V. Babashev, and V. M. Dukelski, Sov.
Phys. JETP \textbf{28}, 404 (1969).

\bibitem{Matveev8} V. B. Matveev and S. V. Babashev, Sov. Phys. JETP \textbf{%
30}, 829 (1970).

\bibitem{Afrosimov2} V. V. Afrosimov, S. V. Bobashev, Yu.S. Gordeev, and V.
M. Lavrov, Sov. Phys. JETP \textbf{35}, 34 (1972).

\bibitem{Francois13} R. Francois, D. Dhuicq, and M. Barat, J. Phys. B: At.
Mol. Phys. \textbf{5,} 963 (1972).

\bibitem{Lorents12} D. C. Lorents and G. M. Conklin J. Phys. B: At. Mol.
Phys. \textbf{5,} 950 (1972).

\bibitem{Latypov11} Z. Z. Latypov, A.A. Shaporenko, Sov. Physics JETP, 
\textbf{42}, 986 (1976).

\bibitem{Kita18} S. Kita, K. Noda, and H. Inouye, J. Chem. Phys. \textbf{63}%
, 4930 (1975).

\bibitem{V. V. Afrosimov16 21} V. V. Afrosimov , Yu .S. Gordeev, and V. M.
Lavrov Sov. Phys. JETP \textbf{41}, 860 (1976).

\bibitem{Bidin6} Yu. F. Bidin and S. S. Godakov, Sov. Phys. JETP Lett. 
\textbf{23}, 518 (1976). \ \ 

\bibitem{Hegerberg3} R. Hegerberg, T. Stefensson, and M. T. Elfort, J. Phys.
B \textbf{11}, 133 (1978).

\bibitem{Jrgensen20} K. Jrgensen, N. Andersen, and J. Olsen, J. Phys. B 
\textbf{11}, 3951 (1978).

\bibitem{Os22} J. O. Olsen, T. Andersen, M. Barat et al., Phys. Rev. A 
\textbf{19}, 1457 (1979).

\bibitem{Kita17} S. Kita, T. Hasegawa, H. Tanuma, and N. Shimakura, Phys.
Rev. A \textbf{52}, 2070 (1995).

\bibitem{Kita2000} S. Kita, S. Gotoh, N. Shimakura, and S. Koseki, Phys.
Rev. 62, 032704 (2000).

\bibitem{Lomsadze23} R. A. Lomsadze, M. R. Gochitashvili, R. V. Kvizhinadze,
N. O. Mosulishvili, and S. V. Bobashev, Tech. Phys. \textbf{52}, 1506 (2007).

\bibitem{Kita10} S. Kita, S. Gotoh, T. Tanaka et al., J. Phys. Soc. Jpn. 
\textbf{76}, 044301 (2007).

\bibitem{KitaR} S. Kita, T. Hatada, and H. Inoue, Y. Shiraishi, and N. Shimakura
J. Phys. Soc. Jpn. \textbf{82}, 124301 (2013).

\bibitem{Kez25} R. A. Lomsadze, M. R. Gochitashvili, R. Ya. Kezerashvili, N.
O. Mosulishvili, and R. Phaneuf, Phys. Rev. A \textbf{87}, 042710 (2013).

\bibitem{Kita19} S. Kita, T. Hattori, and N. Shimakura, J. Phys. Soc. Jpn 
\textbf{84}, 014301 (2015).

\bibitem{Junker} B. R. Junker and J C Browne, Phys. Rev. A \textbf{10,} 2078
(1974).

\bibitem{Sidis} V. Sidis, N. Stolterfoht, and M. Barat, J. Phys. B: At. Mol.
Phys. \textbf{10,} 2815 (1977).

\bibitem{CP} C. R. Christenson et al., Phys. Rev. Lett. \textbf{13}, 138
(1964).

\bibitem{Annu.Rev} E. M. Henley, Annu. Rev. Nucl. Sci. \textbf{19}, 367
(1969).

\bibitem{Blanke} E. Blanke, H. Driller, W. Glockle, H. Genz, A. Richter, and G. Schrieder, Phys. Rev. Lett. 
\textbf{51}, 355 (1983).

\bibitem{Uzikov} A. A. Temerbayev and Yu. N. Uzikov, Phys. Atomic Nucl.,
2015, \textbf{78} (2015).

\bibitem{Alberty} R. A. Alberty, J. Chem. Educ. \textbf{81}, 1206\ (2004).

\bibitem{Fite34} W. L. Fite, R. F. Stebbings, D.G. Hummer, and R.T.
Brackmann, Phys. Rev. \textbf{119}, 663. (1960).

\bibitem{Lavrov35} V. M. Lavrov, R. A. Lomsadze, Abstracts of XIII ICPEAC ,
W. Berlin, p. 610, (1983).

\bibitem{Shingel} R. Shingel and B. H. Branaden, J. Phys. B \textbf{20},
L127 (1987).

\bibitem{Igenbergs37} K. Igenbergs, J. Schweinzer, I. Bray, D. Bridi, and F.
Aumayr, Atomic Data and Nuclear Data Tables \textbf{94}, 981--1014 (2008).

\bibitem{Heer38} F. J. De Heer and T. van Eck, Proc. of III ICPEAC, London,
p.653, (1963).

\bibitem{Kikiani26} B. I. Kikiani, R. A. Lomsadze, and N. O. Mosulishvili et
al., Tech. Phys. \textbf{30}, 934 (1985).

\bibitem{Gochitashvili27} M. R. Gochitashvili, R. A. Lomsadze et al.,
Georgian Electron. Sci. J.: Phys. \textbf{39}-2, (2004).

\bibitem{Gochitashvili28} M. R. Gochitashvili, N. Jaliashvili, R. V.
Kvizhinadze, and B. I. Kikiani, J. Phys. B \textbf{28}, 2453 (1995).

\bibitem{Ajello29} J. M. Ajello and B. Franklin, J. Chem. Phys. \textbf{82},
2519 (1985).

\bibitem{Avakyan30} S. V. Avakyan, R.N. Ii'In, V.M. Lavrov, G.N. Ogurtsov,
Collision processes and excitation of UV emission from planetary atmospheric
gases: a handbook of cross sections, 1999 - books.google.com

\bibitem{Stone31} E. J. Stone and E. C. Zipf, J. Chem. Phys. \textbf{56},
4646 (1972).

\bibitem{Tan32} K. H. Tan, F. C. Donaldson, and J. W. McConkey, Can. J.
Phys. \textbf{52}, 786 (1974).

\bibitem{Tan33} K. H. Tan and J. W. McConkey, Phys. Rev. A \textbf{10}, 1212
(1974).

\bibitem{Barat} M. Barat and W. Lichten, Phys. Rev. A \textbf{6}, 211 (1972).

\bibitem{Solov'ev40} E. A. Solov'ev, Sov. Phys., JETP \textbf{54}, 893
(1981).\ 

\bibitem{Hartree41} R. D. Hartree, The Calculation of Atomic Strictures,
Wiley, New York, 1957.

\bibitem{Bobashev42} S. V. Bobashev and V. A. Kharchenko, Sov. Phys. JETP, 
\textbf{44}, (1976).

\bibitem{Ankudinov43} V. H. Ankudinov, S. V. Bobashev, and V. I. Perel, Sov.
Phys. JETP \textbf{33}, 490 (1971).

\bibitem{Fano24} U. Fano and W. Lichten, Phys. Rev. Lett. \textbf{14}, 627
(1965).

\bibitem{Landau} L. D. Landau, Phys. Z. \textbf{2}, 46 (1932).

\bibitem{Zener} C. Zener, Proc. R. Soc. London A \textbf{137}, 696 (1932).
\end{thebibliography}
\end{document}